\documentclass[10pt,conference]{IEEEtran}
\IEEEoverridecommandlockouts
\usepackage{geometry}
\usepackage{cite}
\usepackage{amsmath,amssymb,amsfonts}
\usepackage{algorithmic}
\usepackage{graphicx}
\usepackage{textcomp}
\usepackage{xcolor}
\def\BibTeX{{\rm B\kern-.05em{\sc i\kern-.025em b}\kern-.08em
    T\kern-.1667em\lower.7ex\hbox{E}\kern-.125emX}}

\DeclareMathOperator*{\argmin}{arg\,min}
\usepackage{mathtools}
\DeclarePairedDelimiter\ceil{\lceil}{\rceil}

\usepackage{authblk}

\DeclareMathOperator{\tr}{tr}

\IEEEoverridecommandlockouts
\IEEEpubid{\makebox[\columnwidth]{978-1-7281-4490-0/20/\$31.00 \copyright~2020~IEEE }
	\hspace{\columnsep}\makebox[\columnwidth]{ }}

\begin{document}

\title{Rate-Splitting Multiple Access for Downlink Multi-Antenna Communications: Physical Layer Design and Link-level Simulations\\
\thanks{This work has been partially supported by the U.K. Engineering and Physical Sciences Research Council (EPSRC) under grant EP/R511547/1, and by Huawei Technologies Co., Ltd.}
}

\author[1]{Onur Dizdar}
\author[1]{Yijie Mao}
\author[2]{Wei Han}
\author[1]{Bruno Clerckx}
\affil[1]{Department of Electical and Electronic Engineering, Imperial College London}
\affil[2]{Huawei Technologies, Shanghai, China}
\affil[ ]{Email: \{o.dizdar,y.mao16,b.clerckx\}@imperial.ac.uk, wayne.hanwei@huawei.com}


\maketitle

\begin{abstract}
Rate-Splitting Multiple Access (RSMA) is an emerging flexible, robust and powerful multiple access scheme for downlink multi-antenna wireless networks. RSMA relies on multi-antenna Rate-Splitting (RS) strategies at the transmitter and Successive Interference Cancellation (SIC) at the receivers, and has the unique ability to partially decode interference and partially treat interference as noise so as to softly bridge the two extremes of fully decoding interference (as in Non-Orthogonal Multiple Access, NOMA) and treating interference as noise (as in Space Division Multiple Access, SDMA or Multi-User Multiple-Input Multiple-Output, MU-MIMO). RSMA has been shown to provide significant room for spectral efficiency, energy efficiency, Quality-of-Service enhancements, robustness to Channel State Information (CSI) imperfections, as well as feedback overhead and complexity reduction, in a wide range of network loads (underloaded and overloaded regimes) and user deployments (with a diversity of channel directions, channel strengths and qualities). RSMA is also deeply rooted and motivated by recent advances in understanding the fundamental limits of multi-antenna networks with imperfect CSI at the Transmitter (CSIT). In this work, we leverage recent results on the optimization of RSMA and design for the first time its physical layer, accounting for modulation, coding (using polar codes), message split, adaptive modulation and coding, and SIC receiver. Link-level evaluations confirm the significant throughput benefits of RSMA over various baselines as SDMA and NOMA. 
\end{abstract}


\section{Introduction}
An essential requirement for the next generation communication systems is to meet the demand for exponentially increasing data rate and accessibility. A key technology expected to meet such requirement is Multi-User Multiple-Input Multiple-Output (MU-MIMO). MU-MIMO is commonly implemented using multi-user linear precoding techniques to separate users' streams at the transmitter. At the receivers, any residual multi-user interference as noise is commonly treated as noise. Such strategy relying on multi-user linear precoding and treat interference as noise at the receivers is also often denoted as Space Division Multiple Access (SDMA).
Though attractive in multi-antenna Gaussian Broadcast Channel (BC) with perfect Channel State Information at the Transmitter (CSIT) for achieving the entire Degree-of-Freedom (DoF) region, SDMA is highly sensitive to CSIT inaccuracy \cite{clerckx_2016}. 

Non-Orthogonal Multiple Access (NOMA) in power-domain is a multiple-access technique based on the well-known Superposition Coding (SC) at the transmitter and Successive Interference Cancellation (SIC) at the receivers. Originally considered for the Single-Input Single-Output (SISO) BC, NOMA has also been studied for multi-antenna scenarios. However, multi-antenna NOMA has been shown to have a significant DoF loss unless the user channels are aligned and have a disparity of strengths \cite{joudeh_2017}. 

Rate-Splitting Multiple Access (RSMA) is a recent multiple-access technique for multi-antenna BC that relies on linearly precoded Rate-Splitting (RS) at the transmitter and SIC at the receivers \cite{clerckx_2016, mao_2018}. RSMA splits the user messages into common and private parts, encodes the common parts of the user messages into a common stream to be decoded by all receivers, encodes the
private parts of the user messages into private streams and
superposes them in a non-orthogonal manner. RSMA has been shown to tackle the problem of imperfect CSIT in multi-antenna BC \cite{clerckx_2016, joudeh_2016_2}.


RSMA manages interference in a flexible and robust manner by partially decoding interference and treating the remaining interference as noise. As a consequence, RSMA has been shown to bridge and outperform existing different multiple access schemes, i.e., SDMA, NOMA, Orthogonal Multiple Access (OMA) and multicasting, and becomes a promising multiple access technique for 5G and beyond \cite{joudeh_2016_2,clerckx_2019,mao_2019}.

Originally introduced for the Single-Input Single-Output Interference Channel (IC), RS has been adapted to multi-antenna BC due to the recent discovery in information theory that RS achieves the entire DoF region of $K$-user Multiple-Input Single-Output (MISO) BC with partial CSIT \cite{piovano_2017,joudeh_2016_2}. Motivated by the DoF benefits of RS, the finite-SNR precoder design of RS in MISO BC has been studied for both perfect CSIT and imperfect CSIT settings under idealistic assumptions of Gaussian signaling and infinite block lengths\cite{joudeh_2017,mao_2018,mao_2018_2,joudeh_2016,joudeh_2016_2,clerckx_2019,mao_2019,piovano_2017}.

In this work, we depart from the existing works on RSMA and relax those assumptions by designing the Physical (PHY)-layer architecture of RSMA with finite constellation modulation schemes, finite length polar codes and Adaptive Modulation and Coding (AMC). We show through Link-Level Simulations (LLS) that RSMA achieves significant throughput gain over SDMA and NOMA.  

\textit{Notation:} Vectors are denoted by bold lowercase letters. 
For any set $\mathcal{S} \subseteq \left\{0,1,\ldots, N-1\right\}$, $\mathcal{S}^{\mathrm{c}}$ denotes its complement. 
For any vector $\mathbf{u}=\left(u_{0}, u_{1},\ldots , u_{N-1}\right)$ and set $\mathcal{S} \subseteq \left\{0,1,\ldots, N-1\right\}$,  \mbox{$\mathbf{u}_{\mathcal{S}}= \left[u_{i} \ : i \in \mathcal{S} \right]$}. The operations $|.|$ and $||.||$ denote the cardinality of a set or absolute value of a scalar and $l_{2}$-norm of a vector, respectively. $\mathbf{a}^{T}$ and $\mathbf{a}^{H}$ denote the transpose and Hermitian transpose of a vector $\mathbf{a}$, respectively. The operation $\otimes$ denotes the Kronecker product and the operation $D^{\otimes}$ denotes Kronecker power of the matrix $D$. $\mathbb{F}^{K}_{2}$ represents the binary field of dimension $K$. $\mathcal{CN}(0,\sigma^{2})$ denotes the Circularly Symmetric Complex Gaussian distribution with zero mean and variance $\sigma^{2}$. $\mathbf{I}$ denotes the identity matrix.
\begin{figure*}[htbp]
	\centerline{\includegraphics[width=6.6in,height=6.6in,keepaspectratio]{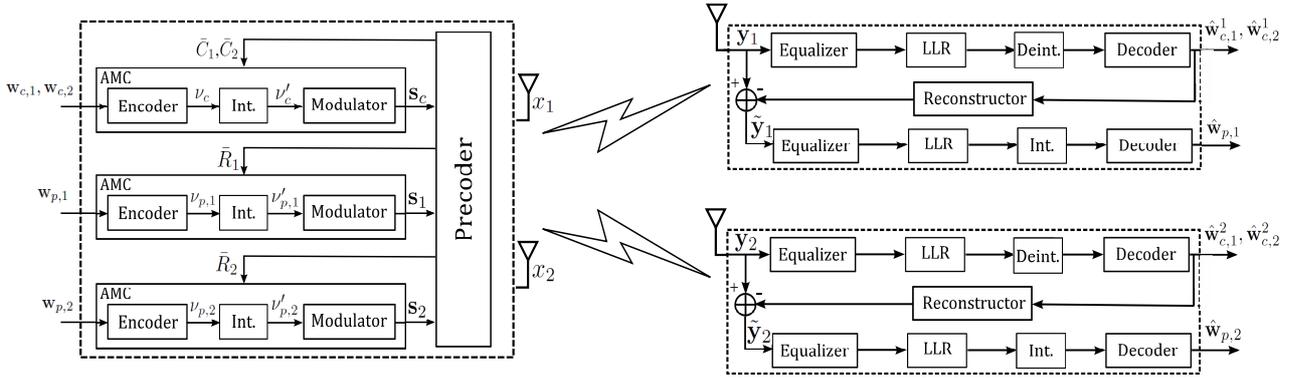}}
	\vspace{-0.15cm}
	\caption{Transmitter and receiver structures}
	\label{fig:system}
	\vspace{-0.45cm}
\end{figure*}
\section{RSMA PHY-Layer Architecture}
\label{sec:phy}
Consider a MISO BC setting consisting of one transmitter with $n_{t}$ transmit antennas and $2$ users with single-antenna each. The messages intended for users $1$ and $2$, $W_{1}$ and $W_{2}$ respectively, are split into common and private parts, i.e., $W_{c,k}$ and $W_{p,k}$, for $k=1,2$. The split messages are assumed to be independent. The common parts of the messages for both users, $W_{c,1}$ and $W_{c,2}$, are combined into the common message $W_{c}$. The common message $W_{c}$ and the private messages are independently encoded into streams $s_{c}$, $s_{1}$ and $s_{2}$, respectively. Linear precoding is applied to all streams with $\mathbf{p}_{c}$, $\mathbf{p}_{1}$ $, \mathbf{p}_{2} \in\mathbb{C}^{n_t}$ being the precoders for the common and private streams, respectively. The resulting transmit signal is 
\begin{align}
	\mathbf{x}=\mathbf{p}_{c}s_{c}+\mathbf{p}_{1}s_{1}+\mathbf{p}_{2}s_{2}.	
\end{align}
We assume that the streams have unit power, so that \mbox{$\mathbb{E}\left\lbrace \mathbf{s}\mathbf{s}^{H}\right\rbrace =\mathbf{I}$} where \mbox{$\mathbf{s}=\left[s_{c}, s_{1}, s_{2}\right]$}. An average transmit power constraint is set as $P_{c}+P_{1}+P_{2} \leq P_{t}$ for $P_{c}=||\mathbf{p}_{c}||^{2}$ and $P_{k}=||\mathbf{p}_{k}||^{2}$, $k=1,2$.

The signal received by user-$k$ is written as
\begin{align}
 \mathbf{y}_{k}=\mathbf{h}_{k}^{H}\mathbf{x}+z_{k}, \quad k=1,2, 
\end{align}
where $\mathbf{h}_{k} \in \mathbb{C}^{n_{t}}$ is the channel vector and $z_{k} \sim \mathcal{CN}(0,1)$ is the Additive White Gaussian Noise (AWGN) component for user-$k$. The receivers apply SIC to detect the common and their corresponding private streams. The common stream is detected first to obtain the common message estimate $\hat{W}_{c}$ by treating the private streams as noise. The common stream is then reconstructed using $\hat{W}_{c}$ and subtracted from the received signal. The remaining signal is used to detect the private messages $\hat{W}_{p,k}$. Finally, the estimated message for user-$k$, $\hat{W}_{k}$, is obtained by combining $\hat{W}_{c,k}$ and $\hat{W}_{p,k}$. 

We consider the general and practical case where the transmitter does not have access to perfect Channel State Information (CSI). We assume a block fading channel model $\mathbf{H}=[\mathbf{h}_{1}, \mathbf{h}_{2}]$ with channel estimation errors, expressed as 
\begin{align}
\mathbf{H}=\sqrt{1-\sigma_{e}^{2}}\widehat{\mathbf{H}}+\sigma_{e}\widetilde{\mathbf{H}},
\end{align}
where $\widehat{\mathbf{H}}=[\widehat{\mathbf{h}}_{1}, \widehat{\mathbf{h}}_{2}]$ is the channel estimate at the transmitter with independent and identically distributed (i.i.d.) elements $\widehat{h}_{j,k}\sim\mathcal{CN}(0,1)$, $\widetilde{\mathbf{H}}=[\widetilde{\mathbf{h}}_{1}, \widetilde{\mathbf{h}}_{2}]$ is the channel estimation error with i.i.d. elements $\widetilde{h}_{j,k}\sim\mathcal{CN}(0,1)$, for $1\leq j \leq n_{t}$ and, $\widehat{h}_{j,k}$ and $\widetilde{h}_{j,k}$ are the elements of the $j$th row and $k$th column of $\widehat{\mathbf{H}}$ and $\widetilde{\mathbf{H}}$, respectively, for \mbox{$k=1,2$}. The error variance is modeled as $\sigma_{e}^{2}=P^{-\alpha}$, where $\alpha$ is named as the CSIT quality scaling factor. We assume perfect CSI at the receivers.

The instantaneous rates at user-$k$ for the common and the $k$th private
stream are written as \mbox{$R_{c,k}=\log_{2}(1+\gamma_{c,k})$} and \mbox{$R_{k}=\log_{2}(1+\gamma_{k})$}, where $\gamma_{c,k}$ and $\gamma_{k}$ are the Signal-to-Interference-plus-Noise Ratio (SINR) levels for the common and private streams, respectively. 
We define ergodic rates (ERs) as \mbox{$E_{\mathbf{H}}\left\lbrace R_{c,k}\right\rbrace$} and \mbox{$E_{\mathbf{H}}\left\lbrace R_{k}\right\rbrace$}, and the average rates (ARs) as \mbox{$\bar{R}_{c,k}=E_{\mathbf{H}|\widehat{\mathbf{H}}}\left\lbrace R_{c,k}|\widehat{\mathbf{H}}\right\rbrace$} and  \mbox{$\bar{R}_{k}=E_{\mathbf{H}|\widehat{\mathbf{H}}}\left\lbrace R_{k}|\widehat{\mathbf{H}}\right\rbrace$}. It is shown in \cite{joudeh_2016_2} that the ERs can be characterized by averaging the ARs over the variation in $\widehat{\mathbf{H}}$. The AR for the common stream is set as \mbox{$\bar{R}_{c}=\min(\bar{R}_{c,1},\bar{R}_{c,2})$.}

We consider the precoding scheme in \cite{joudeh_2016_2}, in which the objective is to maximize the ergodic sum-rate (ESR) under imperfect CSIT. 
The optimized precoders are the solutions of the equivalent average sum-rate (ASR) maximization problem
\begin{subequations}
\begin{alignat}{3}
\max_{\bar{C}_{1}, \bar{C}_{2}, \mathbf{P}}&     \quad  \sum_{k=1}^{2}(\bar{C}_{k}+\bar{R}_{k})          \label{eqn:obj}   \\
\text{s.t.}&  \quad  \bar{R}_{c,k}    \geq \bar{C}_{1}+\bar{C}_{2}, \quad  \quad &k =1,2 \label{eqn:common_rate_1} \\
& \quad\bar{C}_{k} \geq 0, \quad  &k =1,2 \label{eqn:common_rate_2} \\
& \quad\bar{C}_{k}+\bar{R}_{k} \geq R_{0}, \quad  &k =1,2 \label{eqn:total_rate} \\
& \quad    \tr(\mathbf{P}\mathbf{P}^ {H}) \leq P_{t},  \label{eqn:avg_power}
\end{alignat}
\label{eqn:asr_problem}
\end{subequations}
\hspace{-0.1cm}where \mbox{$\mathbf{P}=[\mathbf{p}_{c}, \mathbf{p}_{1}, \mathbf{p}_{2}]$} is the precoder matrix for the common and private streams. $\bar{C}_k$ is the part of the AR for the common message allocated for the transmission of $W_{c,k}$, for $k=1,2$. As the message of each user is transmitted through two streams, the total AR of each user is equal to $\bar{C}_k+\bar{R}_k$, which is the sum of the ARs of the common message and private message for user-k. Constraint \eqref{eqn:common_rate_1} ensures that the common stream is successfully decoded by both users. Constraint \eqref{eqn:total_rate} ensures the Quality-of-Service (QoS) rate of each user and $R_0$ is the minimum rate of each user. Constraint \eqref{eqn:avg_power} is the transmit power constraint at the transmitter.
Note that RSMA boils down to SDMA by switching off the precoder of the common stream and encoding the user messages only into the private streams $s_{1}, s_{2}$, and to NOMA by switching off the precoder of the private stream for the weak user and encoding its message into the common stream $s_{c}$ \cite{joudeh_2016, mao_2018, clerckx_2019}. 

The stochastic problem in \eqref{eqn:asr_problem} is first transformed into an approximate deterministic form using the Sample Average Approximation (SAA) method \cite{shapiro_2009}. The approximate deterministic form is transformed into an equivalent Augmented Weighted Sum Mean Square Error problem, which can be solved by the Alternating Optimization algorithm. We omit the details of the applied transformations and algorithm due to lack of space. Interested reader can refer to \cite{joudeh_2016_2} for details.


We consider the transmitter and receiver architectures shown in Fig.~\ref{fig:system} for RSMA. Detailed explanations for the modules are given in the following subsections.
\subsection{Transmitter}
The common and private messages are mapped to bit vectors $\mathbf{w}_{c,1}, \mathbf{w}_{c,2}, \mathbf{w}_{p,1}$ and $\mathbf{w}_{p,2}$ of lengths $K_{c,1}, K_{c,2}, K_{p,1}$ and $K_{p,2}$, respectively. 
The common and private information bit vectors are independent and uniformly distributed in $\mathbb{F}_{2}^{K_{c,i}}$ and $\mathbb{F}_{2}^{K_{p,i}}$, $i=1,2$, respectively, based on the independence assumption of the split messages.
The information bit vectors corresponding to the common messages are jointly encoded into the common coded vector $\mathbf{\nu}_{c} \in \mathbb{F}^{N_{c}}_{2}$, where $N_{c}$ is the code block length of the common coded vector. The information bit vectors corresponding to the private messages are encoded separately into the private coded vectors $\mathbf{\nu}_{p,1} \in \mathbb{F}^{N_{p,1}}_{2}$ and $\mathbf{\nu}_{p,2} \in \mathbb{F}^{N_{p,2}}_{2}$, where $N_{p,i}$, $i=1,2$, are the code block lengths of the private coded vectors. The coded bit vectors are then modulated separately into common and private streams $\mathbf{s}_{c},\mathbf{s}_{1}$, and $\mathbf{s}_{2}$ respectively, each of length $S$. 

The modulation and coding parameters for  $\mathbf{s}_{c},\mathbf{s}_{1}$, and $\mathbf{s}_{2}$ are determined by an AMC algorithm according to the ARs.
Interleavers are used between the encoders and modulators in accordance with the Bit-Interleaved Coded Modulation (BICM) scheme \cite{caire_1998}. 
The details of encoding, modulation, and AMC algorithm are given in the following subsections. 
\subsubsection{Encoder}
\label{sec:encoder}
The common stream is decoded by treating the private messages as noise and its rate is set as \mbox{$\bar{R}_{c}=\min(\bar{R}_{c,1},\bar{R}_{c,2})$} so as to ensure the successful decoding of $s_{c}$ at both receivers.
The problem of achieving such rate is described as the max-min problem for BC in \cite{cover_1972}.
A point-to-point coding scheme used with a proper modulation and code rate to transmit at rates approaching $\bar{R}_{c}$ in compound settings is suitable for such transmission.

We consider polar coding in the system. Aside from its good error correction capabilities with advanced decoding methods (e.g., Successive Cancellation List, SCL), a major advantage of polar coding is its inherent code design flexibility with a fine granularity in code rate, which is useful for the AMC algorithm (explained in Section~\ref{sec:AMC}). 

Polar codes are a class of linear block codes that are proved to achieve the symmetric capacity of any Binary-Input Discrete Memoryless Channel (B-DMC) \cite{arikan_2009}. The block length of a polar code is represented by $N=2^{n}$, where $n$ is an integer and $n>0$. The polar encoding operation can be expressed as $\mathbf{\nu}=\mathbf{u}\mathrm{G}. $
The encoding matrix $\mathrm{G}$ is written as $\mathrm{G} = \mathrm{B}_{N}\mathrm{F}^{\otimes n}$, where $\mathrm{F} = \begin{bmatrix}		1 & 0  \\		1 & 1 \end{bmatrix}  \nonumber $ and $\mathrm{B}_{N}$ is the bit-reversal matrix. 
The vector $\mathbf{u} \in \mathbb{F}_{2}^{N}$ is the uncoded bit vector consisting of $K$ information bits and $N-K$ redundant bits. 
The locations of information bits in the vector $\mathbf{u}$ are chosen from the set $\mathcal{A}$, which is the set of indexes of $K$ polarized channels with smallest Bhattacharyya parameters. The binary elements of the information bit vectors $\mathbf{w}_{c,1}, \mathbf{w}_{c,2}, \mathbf{w}_{p,1}$ and $\mathbf{w}_{p,2}$ are placed into the bit locations reserved for information bits in the corresponding uncoded bit vectors $\mathbf{u}_{c}$, $\mathbf{u}_{1}$, and $\mathbf{u}_{2}$.
The locations of redundant bits, which are named as \textsl{frozen bits}, are chosen from the set $\mathcal{A}^{\mathrm{c}}$, where \mbox{$\mathcal{A} \cap \mathcal{A}^{\mathrm{c}}=\O$} and \mbox{$\mathcal{A}\cup \mathcal{A}^{\mathrm{c}}=\left\{0, 1, \ldots, N-1\right\}$}. 
The locations and values of frozen bits are known by the encoder and decoder. 

Denote the sets of information bit indexes for the common and private streams as $\mathcal{A}_{c}$, $\mathcal{A}_{1}$, and $\mathcal{A}_{2}$. Denote the vectors of information bits in the uncoded bit vectors of the common and private messages, $\mathbf{u}_{c}$, $\mathbf{u}_{1}$, and $\mathbf{u}_{2}$, as $\mathbf{u}_{c,\mathcal{A}_{c}}$, $\mathbf{u}_{1,\mathcal{A}_{1}}$ and $\mathbf{u}_{2,\mathcal{A}_{2}}$, respectively. The private information bit vectors
are placed into the information bit locations by setting \mbox{$\mathbf{u}_{k,\mathcal{A}_{k}}=\mathbf{w}_{p,k}$}, for $k=1,2$. The joint encoding of the common information bit vectors of the two users, which are independently generated, is performed by appending the vectors to be assigned to the information bit locations, i.e., $\mathbf{w}_{c}=\left[\mathbf{w}_{c,1}, \mathbf{w}_{c,2} \right]$. 
Since the common information bit vectors of both users need to be decoded correctly at the receivers, all common information bits have equal importance to each user. Thus, no specific rule is considered for placing the common information bits in the uncoded bit vector, which is set as $\mathbf{u}_{c,\mathcal{A}_{c}}=\mathbf{w}_{c}$.

Since the common message is intended for both users, $\mathcal{A}_{c}$ should be designed for the compound setting consisting of two channels. Denoting the information sets for two channels as $\mathcal{A}_{1}$ and $\mathcal{A}_{2}$, a possible approach is to set $\mathcal{A}_{c}=\mathcal{A}_{1} \cap \mathcal{A}_{2}$, which falls short of the compound capacity of the two channels. 
A chaining operation is described in \cite{hassani_2014} to design polar codes for such compound settings. This approach aligns the information sets $\mathcal{A}_{1}$ and $\mathcal{A}_{2}$ corresponding to the two different channels by chaining over $k$ separate blocks of polar codes, so that the rate of the construction, $
\lim_{N\to\infty}\frac{|\mathcal{A}_{1} \cap \mathcal{A}_{2}|+\frac{k-1}{k}|\mathcal{A}_{1}\backslash\mathcal{A}_{2}|}{N}, 
$
approaches the compound channel capacity as $k$ tends to infinity.
In practice, such approach is more suitable for scenarios with perfect CSIT, as the information sets $\mathcal{A}_{1}$ and $\mathcal{A}_{2}$ determined using the imperfect CSIT may be different from those for the instantaneous channel realizations.
Therefore, we focus on minimizing the average error probability for code construction in order to increase the robustness of the system to imperfect CSIT
\footnote{Another approach is to construct codes for throughput maximization \cite{khoshnevis_2018}. However, such approach is more suitable for scenarios with perfect CSIT.} 
and use the Gaussian approximation method \cite{trifonov_2012}, \cite{li_2013}. We use Cyclic Redundancy Check (CRC) codes as outer codes for both private and common messages, in order to enhance the error performance of the polar codes \cite{tal_vardy_2015}.

\subsubsection{Modulator}
We consider Quadrature Amplitude Modulation (QAM) modulation schemes, specifically $4$-QAM, $16$-QAM, $64$-QAM and $256$-QAM. We use the shortening method in \cite{li_2015} to adjust the code block length for $64$-QAM, as the block lengths of polar codes are in the form $N=2^{n}$, $n>0$. 

Consider a modulation scheme with alphabet $\mathcal{M}$ and modulation order $|\mathcal{M}|=2^{m}$. The consecutive encoded and interleaved bits $(\nu^{\prime}_{mi+0}, \nu^{\prime}_{mi+1}, \ldots, \nu^{\prime}_{mi+m-1})$, \mbox{$i \in \left\lbrace 0,1,\ldots,N\backslash m\right\rbrace $} are mapped to a constellation signal $s \in \mathcal{M}$ according to a labeling rule \mbox{$
 \mathcal{L} : \left\lbrace 0,1\right\rbrace ^{m} \rightarrow \mathcal{M}$},
which is chosen as the Gray labeling in the proposed architecture.
\footnote{One can consider other bit mappings for BICM with polar codes, e.g. \cite{seidl_2013}. We choose the conventional Gray mapping to preserve the generic structure. }
\subsubsection{Precoder}
\label{sec:precoding}
The employed precoders are the solutions of the maximization problem in \eqref{eqn:asr_problem}. In addition to the transmitter beamforming, the obtained precoders are used to calculate the ARs for the AMC algorithm. Using SAA, the SINR values for the common and private streams are calculated by taking the
ensemble averages over $M$ realizations of the channel, i.e.,  
\begin{subequations}
\begin{align}
\bar{R}_{c,k}&=\frac{1}{M}\sum_{m=1}^{M}\log_{2}\left( 1+\frac{|(\mathbf{h}_{k}^{(m)})^{H}\mathbf{p}_{c}|^{2}}{\sum_{i=1}^{2}|(\mathbf{h}_{k}^{(m)})^{H}\mathbf{p}_{i}|^{2}+1}\right) , \\
\bar{R}_{k}&=\frac{1}{M}\sum_{m=1}^{M}\log_{2}\left( 1+\frac{|(\mathbf{h}_{k}^{(m)})^{H}\mathbf{p}_{k}|^{2}}{|(\mathbf{h}_{k}^{(m)})^{H}\mathbf{p}_{i}|^{2}+1}\right) ,  i\neq k, 
\end{align} 
\label{eqn:ARs}
\end{subequations}
\hspace{-0.1cm}where $\mathbf{h}_{k}^{(m)}$ represents the $m$th channel realization for user-$k$ for a given channel estimate $\widehat{\mathbf{h}}_{k}$. 
\subsubsection{AMC Algorithm}	
\label{sec:AMC}
The aim of AMC is to maximize the system throughput by selecting appropriate Modulation and Coding Scheme (MCS) for transmission depending on the channel characteristics. 
Throughput is a function of the amount of information carried by the selected MCS and the error performance of the transmissions. A formulation of throughput for a given MCS can be written as $T=R(1-P_{B})$, where $R$ is the spectral efficiency of the MCS and $P_{B}$ is the Block Error Rate (BLER) of the MCS in a given setting.

Designing an AMC algorithm involves estimating the link quality and choosing an appropriate MCS for the estimated link quality. A conventional metric for the link quality is the link Signal-to-Noise Ratio (SNR) or SINR. 
Since throughput is a function of rate and BLER, choosing an appropriate MCS involves predicting the achievable BLER by the target system for any given SNR/SINR and channel realization. The estimated link quality may not predict the actual BLER performance of the system accurately due to several factors, such as finite length coding, imperfect CSIT, performance of employed coded modulation scheme, and suboptimal detection and decoding methods at the receiver. As a result, the achieved BLER may be higher than the estimated one. 

Several approaches have been suggested to tackle the problem mentioned above.
One method is to put a margin from the estimated SINR and choose an MCS with lower spectral efficiency than the one corresponding to the estimated link quality. The imperfections in the channel estimates are considered in the SINR calculations to design AMC algorithms with good performance under imperfect CSIT \cite{zhang_2018_1, ormeci_2001, zhou_2004}.

We propose an AMC algorithm for the transmitter architecture. 
The algorithm makes use of the ARs $\bar{R}_{c}$, $\bar{R}_{1}$, and $\bar{R}_{2}$ as metrics for the link qualities. 
The proposed AMC algorithm takes into account the effects of imperfect CSIT inherently due to the ARs being calculated over $M$ channel ensembles as in \eqref{eqn:ARs}.
As opposed to the algorithms in the literature, we do not pick the MCSs from a table of predetermined modulation schemes and code rates. Instead, we first determine a modulation scheme (from a predetermined set) for a given AR and obtain a code rate that achieves the closest possible transmission rate to the corresponding AR, owing to the flexible nature of polar codes in terms of code rates.

We define the modulation alphabets $\mathcal{M}_{c}$, $\mathcal{M}_{1}$, $\mathcal{M}_{2}$ 
and code rates $r_{c}$, $r_{1}$, $r_{2}$ for the common and private streams, respectively. 
We also define the set of all modulation alphabets used in the system as $\mathcal{Q}$.
The set of feasible modulation schemes for a given rate $\bar{R}_{k}$ is found as
\begin{align}
\mbox{$\mathcal{Q}(\bar{R}_{k},\beta)\hspace*{-0.05cm}=\hspace*{-0.05cm}\left\lbrace \mathcal{M}: \log_{2}(|\mathcal{M}|) \geq \min\left( \frac{\bar{R}_{k}}{\beta }, M^{\dagger}\right) , \mathcal{M} \in \mathcal{Q} \right\rbrace $ }\hspace*{-0.15cm}, \nonumber
\end{align} 
where $\beta$ is the maximum code rate and $M^{\dagger}$ is the logarithm of the highest modulation order used in the architecture. The parameter $\beta$ can be determined based on the performance of the employed modulation schemes, channel coding method and the coded modulation scheme. 
The modulation schemes for the common and private messages are determined as
\begin{align}
\mathcal{M}_{l}=\argmin_{\mathcal{M}\in\mathcal{Q}(\bar{R}_{l},\beta)}(|\mathcal{M}|), \quad \quad l \in \left\lbrace c,1,2\right\rbrace. 
\end{align}

The coded block lengths are calculated by \mbox{$N_{l}=S\log _{2}(|\mathcal{M}_{l}|)$}, $l \in \left\lbrace c,1,2\right\rbrace$, for a given stream length $S$. Finally, the code rates are calculated as 
\begin{align}
 r_{l} &= \ceil{N_{l}\min(\bar{R}_{l}/|\mathcal{M}_{l}|,\beta)}/N_{l}, \quad \quad l \in \left\lbrace c,1,2\right\rbrace. 
\end{align}
  
In order to compensate for the performance loss due to finite-length channel coding, the BICM scheme, the employed bit mapping, and imperfect CSIT, we perform energy back-off from the rates given in \eqref{eqn:ARs}. 
\begin{figure}[t!]
	\centerline{\includegraphics[width=3.4in,height=3.4in,keepaspectratio]{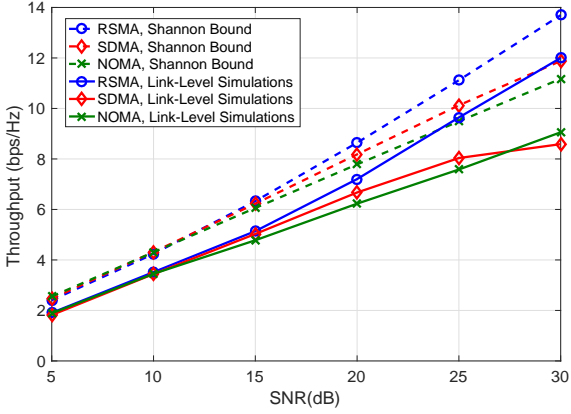}}
	\vspace{-0.15cm}
	\caption{SNR vs. Throughput, $\alpha=0.6$, No QoS constraint}
	\label{fig:tp_alpha06_noQoS}
	\vspace{-0.45cm}
\end{figure}
\subsection{Receiver}
The received signal is first detected using an equalizer. Log-Likelihood Ratios (LLRs) are calculated from the equalized signal for Soft Decision (SD) decoding of polar codes. 
The receiver employs a Hard Decision (HD) SIC algorithm, which performs signal reconstruction from the HDs for the decoded bits at the output of the polar decoder. 
The reconstruction is performed by replicating the operations at the transmitter for the decoded bits to obtain a precoded signal and multiplying the resulting signal with the user channel vector.
\subsubsection{Equalizer}							
We employ Minimum Mean Square Error (MMSE) equalizers for the common and private streams. The MMSE equalizers $g^{\prime}_{c,k}$ for the common streams are calculated by minimizing \mbox{$E\left\lbrace|g_{c,k}y_{k}-s_{c} |^{2}\right\rbrace$}, so that
\mbox{$g^{\prime}_{c,k}=\mathbf{p}_{c}^{H}\mathbf{h}_{k}/(|\mathbf{h}_{k}^{H}\mathbf{p}_{c}|^{2}+\sum_{j=1}^{2}|\mathbf{h}_{k}^{H}\mathbf{p}_{j}|^{2}+1)$,
for} \mbox{$k=1,2$}. Similarly, the MMSE equalizers $g^{\prime}_{k}$ for the private streams are calculated by minimizing \mbox{$E\left\lbrace|g_{k}\tilde{y}_{k}-s_{k} |^{2}\right\rbrace $} so that
\mbox{$ g^{\prime}_{k}=\mathbf{p}_{k}^{H}\mathbf{h}_{k}/(\sum_{j=1}^{2}|\mathbf{h}_{k}^{H}\mathbf{p}_{j}|^{2}+1)$
for} \mbox{$k=1,2$}, where \mbox{$\tilde{y}_{k}=y_{k}-\mathbf{h}_{k}^{H}\mathbf{p}_{c}s_{c}$} is the received signal after the reconstructed common stream is cancelled.
\subsubsection{LLR Calculator}
We use the LLR calculation method in \cite{seethaler_2004}. Denote the LLR of the $i$-th bit in the $l$-th symbol of the common stream for the $k$th user by $\lambda_{k,l}^{(i)}$. We write 
\begin{align}
\lambda_{c,k,l}^{(i)}=\gamma_{c,k}\left[\min_{a\in \theta_{1}^{(i)}}\psi(a)-\hspace{-0.2cm}\min_{a\in \theta_{0}^{(i)}}\psi(a) \right], \nonumber
\end{align}
\hspace*{-0.05cm}where $\theta_{b}^{(i)}$ is the set of modulation symbols with the value $b$, $b \in {0,1}$ at the $i$-th bit location, \mbox{$\psi(a)=|\frac{g^{\prime}_{c,k}y_{k}}{\rho_{c,k}}\hspace{-0.1cm}-a|^{2}$}, and $\rho_{c,k}=\gamma_{c,k}/(1+\gamma_{c,k})$. The LLRs of the bits in private streams are calculated in a similar manner, with $\gamma_{c,k}$, $\rho_{c,k}$, $g^{\prime}_{c,k}$, and $y_{k}$ replaced with $\gamma_{k}$, $\rho_{k}$ $g^{\prime}_{k}$ and $\tilde{y}_{k}$, respectively.

\subsubsection{Decoder}
As explained in Section~\ref{sec:encoder}, using point-to-point coding methods suffice to satisfy the common rate. Thus, we employ a conventional CRC-aided SCL decoder \cite{tal_vardy_2015}. 
\section{Link-Level Simulation Results}
\label{sec:results}
We demonstrate the performance improvements achieved by RSMA over SDMA and NOMA by LLS. We consider a MISO BC scenario in which a central node employing $n_{t}=2$ antennas aims to communicate with $2$ users.
\begin{figure}[t!]
	\centerline{\includegraphics[width=3.4in,height=3.4in,keepaspectratio]{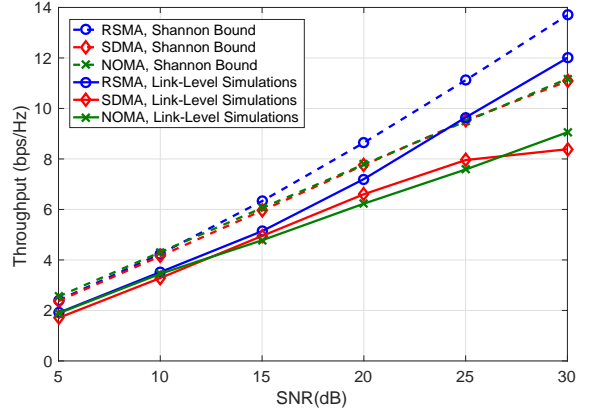}}
	\vspace{-0.15cm}
	\caption{SNR vs. Throughput, $\alpha=0.6$, $R_{0}=0.1$bps/Hz}
	\label{fig:tp_alpha06_QoS}
	\vspace{-0.45cm}
\end{figure}

We investigate the throughput levels achieved by RSMA, SDMA and NOMA for different average received SNR levels. Let $S^{(l)}$ denote the number of channel uses in the $l$-th Monte-Carlo realization and $D_{s,k}^{(l)}$ denote the number of successfully recovered information bits by user-$k$ in the common stream (excluding the part of the common message intended for the other user) and its private stream. We obtain the throughput as  
\begin{align}
\mathrm{Throughput[bps/Hz]}=\frac{\sum_{l}(D_{s,1}^{(l)}+D_{s,2}^{(l)})}{\sum_{l}S^{(l)}}.
\end{align}

We fix the modulated block length $S^{(l)}=256$ throughout the simulations.
The maximum code rate is set as $\beta=0.9$. The energy back-off values for the AMC scheme described in Section~\ref{sec:AMC} are chosen to maximize the throughput at a given SNR while satisfying BLER $\leq10^{-1}$ simultaneously and their values are obtained by simulations. 

Figures~\ref{fig:tp_alpha06_noQoS}~and~\ref{fig:tp_alpha06_QoS} show the Shannon bounds and throughput levels achieved by RSMA, SDMA, and NOMA for $\alpha=0.6$ with and without QoS constraints, respectively. The QoS constraint in Fig.~\ref{fig:tp_alpha06_QoS} is set as $R_{0}=0.1$~bps/Hz. Assigning higher rates to $R_{0}$ frequently results in infeasible problems for SDMA for the considered settings.

An immediate observation from the figures is the matching trends of the Shannon bounds and  throughput curves obtained by LLS. 
The performance results clearly demonstrate that RSMA achieves a significant throughput gain over SDMA and NOMA. Comparing the figures, one can observe that the performances of RSMA and NOMA are not affected by the QoS constraint, while the performance of SDMA degrades, even under a QoS constraint of $0.1$~bps/Hz. As mentioned above, higher QoS constraint rates result in infeasible problems for SDMA, implying that SDMA will not be able to support multiple users under such constraints.  

\begin{figure}[t!]
	\centerline{\includegraphics[width=3.4in,height=3.4in,keepaspectratio]{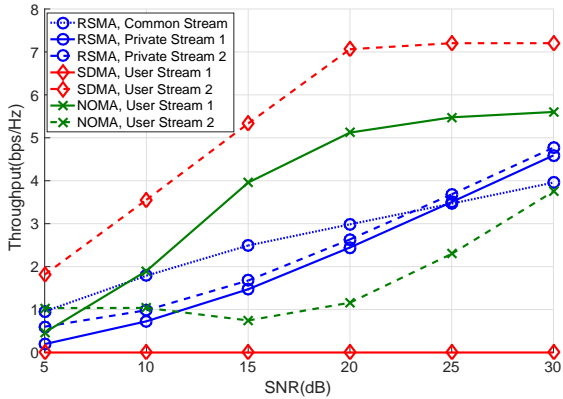}}
	\vspace{-0.15cm}
	\caption{SNR vs. Throughput of each stream, $\alpha=0.6$}
	\label{fig:streams}
	\vspace{-0.45cm}
\end{figure}

During the Monte Carlo simulations with different channel realizations, it was observed that SDMA switches to single-user transmission (i.e. OMA) frequently. This results in transmission rates which exceed the maximum spectral efficiency achievable by the largest modulation order and code rate ($\log_{2}(|\mathcal{M}_{256-\mathrm{QAM}}|)\beta=7.2$bps/Hz) to be assigned to a single stream. Figure~\ref{fig:streams} demonstrates this phenomenon for a given realization.
As seen from the figure, SDMA assigns the transmit power to a single stream in order to maximize the sum-rate under interference, whereas RSMA and NOMA distribute the power among multiple streams. The throughput saturates when the transmission rate exceeds $7.2$bps/Hz.  
The saturated throughput performance of SDMA in Figures~\ref{fig:tp_alpha06_noQoS}~and~\ref{fig:tp_alpha06_QoS} at the high SNR region shows that the frequency of single-user transmission (or transmission by assigning almost all transmit power to one user when QoS constraints are enforced) is considerably high in SDMA. 

Another result of the phenomenon explained above is observed through the throughput performances of SDMA with and without QoS constraints. One can notice from the Figures~\ref{fig:tp_alpha06_noQoS}~and~\ref{fig:tp_alpha06_QoS} that the Shannon bound for SDMA with QoS constraints is degraded with respect to the Shannon bound for SDMA without the QoS constraints. However, such degradation is not observed in the SDMA throughput curves for the two cases (with and without QoS constraints), which is due to the saturated throughput of SDMA for both cases. 

\section{Conclusion}
In this work, we analyzed the performances of RSMA, SDMA and NOMA by LLS under imperfect CSIT. We design basic transmitter and receiver architectures for RSMA with finite constellation modulation schemes, polar codes and an AMC algorithm. Using the proposed architectures and our LLS platform, we demonstrate that RSMA is more robust and achieves significantly higher throughput than SDMA and NOMA with and without QoS constraints in the considered imperfect CSIT settings.

\end{document}